\documentclass{jaa}
\usepackage{natbib}
\usepackage{graphicx}

\begin{document}\sloppy

\title{Science with the {\em AstroSat} Soft X-ray Telescope: an overview}


\author{Sudip Bhattacharyya\textsuperscript{1,*}, Kulinder Pal Singh\textsuperscript{2}, Gordon Stewart \textsuperscript{3}, Sunil Chandra\textsuperscript{4}, Gulab C. Dewangan\textsuperscript{5}, Nilima S. Kamble\textsuperscript{1}, Sandeep Vishwakarma\textsuperscript{1}, Jayprakash G. Koyande\textsuperscript{1} and Varsha Chitnis\textsuperscript{6}}
\affilOne{\textsuperscript{1}Department of Astronomy and Astrophysics, Tata Institute of Fundamental Research, 1 Homi Bhabha Road, Mumbai 400005, India.\\}
\affilTwo{\textsuperscript{2}Indian institute of Science Education and Research Mohali, Sector 81, SAS Nagar, Manauli PO, 140306, India.\\}
\affilThree{\textsuperscript{3}Department of Physics and Astronomy, The University of Leicester, University Road, Leicester LE1 7RH, UK.\\}
\affilFour{\textsuperscript{4}Centre for Space Research, North-West University, Potchefstroom 2520, South Africa.\\}
\affilFive{\textsuperscript{5}Inter-University Centre for Astronomy and Astrophysics (IUCAA), PB No. 4, Ganeshkhind, Pune 411007, India.\\}
\affilSix{\textsuperscript{6}Department of High Energy Physics, Tata Institute of Fundamental Research, 1 Homi Bhabha Road, Mumbai 400005, India.\\}



\twocolumn[{

\maketitle

\corres{sudip@tifr.res.in}

\msinfo{06 November 2020}{04 December 2020}

\begin{abstract}
	The Soft X-ray Telescope (SXT) aboard the {\it AstroSat} satellite is the first Indian X-ray telescope in space. It is a modest size X-ray telescope with a charge coupled device (CCD) camera in the focal plane, which provides X-ray images in the $\sim 0.3-8.0$~keV band. A forte of SXT is in providing undistorted spectra of relatively bright X-ray sources, in which it excels some current large CCD-based X-ray telescopes. Here, we highlight some of the published spectral and timing results obtained using the SXT data to demonstrate the capabilities and overall performance of this telescope.
\end{abstract}

\keywords{galaxies: active---novae, cataclysmic variables---space vehicles: instruments
---telescopes---X-rays: binaries---X-rays: stars.}

}]


\doinum{12.3456/s78910-011-012-3}
\artcitid{\#\#\#\#}
\volnum{000}
\year{0000}
\pgrange{1--}
\setcounter{page}{1}
\lp{1}

\section{Introduction}\label{Introduction}

The {\it AstroSat} observatory is the first dedicated Indian astronomy satellite
\citep{Agrawal2006,Singhetal2014,SeethaMegala2017,SinghBhattacharya2017}, 
with the Soft X-ray Telescope 
(SXT\footnote{https://www.tifr.res.in/$\sim$astrosat\_sxt/index.html}), 
which is the first Indian X-ray telescope in space, on board
\citep{Singhetal2016,Singhetal2017a}.

Cosmic X-rays are reflected by two sets of co-axial nested mirrors in SXT.
The first set has conically approximated paraboloid surfaces and
the second set has conically approximated hyperboloid surfaces.
This is an approximate Wolter I geometry \citep{Wolter1952}, with a cooled 
charge coupled device (CCD) camera at the focal plane.
Multiple instruments/telescopes of {\it AstroSat} can simultaneously
observe a source in a wide energy range from optical to hard X-rays 
\citep[up to $\sim 100 $~keV; ][]{SeethaMegala2017}, and 
SXT covers the crucial soft X-ray band \citep[$\sim 0.3-8.0$~keV; ][]{Singhetal2017b} 
in this range. This telescope has the following modest capabilities
\citep{SinghBhattacharya2017,Singhetal2017a,Singhetal2017b}:
(1) the maximum effective area of $\sim 90$~cm$^2$ at $\sim 1.5$~keV;
(2) an energy resolution of $80-150$ eV in the $0.3-8.0$ keV range;
(3) time resolutions of 2.37 s in the Photon Counting (PC) mode and
of 0.278 s in the Fast Window (FW) mode;
(4) the field of view of 40 arcmin square; and
(5) the Point Spread Function (PSF) having a Full Width Half Maximum 
of $\sim 100$ arcsec and the half encircled energy radius of $\sim 5.5$ arcmin.
However, SXT has a much smaller pile-up compared to current large soft X-ray 
imaging telescopes, and hence is ideal to observe bright X-ray point sources.

These make SXT capable, as an independent telescope, to study continuum spectra, 
broad and somewhat narrow spectral lines and variations with timescales of 
seconds and above of various types of cosmic sources. Furthermore, this telescope, 
jointly with the Large Area X-ray Proportional Counters (LAXPC) and the
Cadmium-Zinc-Telluride Imager (CZTI) aboard {\it AstroSat}, or jointly with
any other hard X-ray instrument or telescope, such as {\it NuSTAR},
can observe the broadband X-ray spectrum of cosmic sources, and can
uniquely contribute to the estimation of the hydrogen column density,
and the characterization of the soft X-ray spectra.
Its good energy resolution and signal-to-noise ratio can be particularly
useful for broadband spectral modeling.

\begin{figure*}[t]
\centering\includegraphics[height=.35\textheight,angle=-90]{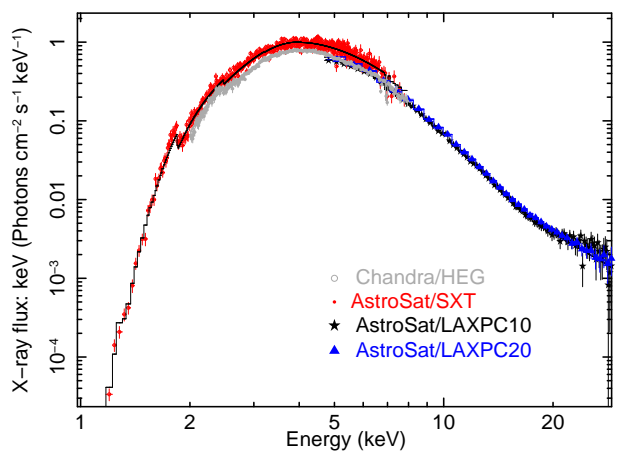}
	\caption{Joint fitting of X-ray spectra of the BHXB 4U 1630--47 from {\it AstroSat} SXT, two LAXPC detectors (LAXPC10 and LAXPC20) and {\it Chandra} High Energy Grating (HEG). An absorbed, relativistic, disk blackbody model with Gaussian absorption features and convolved with a Comptonization model is used for fitting. The source was in a soft state with a disk flux fraction of $\sim 0.97$. It can be seen that SXT covers the spectrum and its peak better than other instruments, which was particularly useful to measure the black hole spin \citep[see section~\ref{BHXB}; figure courtesy: Mayukh Pahari; ][]{Paharietal2018a}.
\label{4u1630-spec-sudip-v3.ps}}
\end{figure*}

In this paper, we give an overview of some notable scientific results 
which required a significant role of SXT. In sections
\ref{BHXB}, \ref{NSLMXB}, \ref{NSHMXB}, \ref{ULP}, \ref{CV}, 
\ref{AGN} and \ref{star}, we mention results on 
black hole X-ray binaries (BHXBs), neutron star low-mass X-ray binaries
(LMXBs), neutron star high-mass X-ray binaries (HMXBs),
ultra-luminous X-ray pulsars (ULPs), cataclysmic variables (CVs),
active galactic nuclei (AGNs) and stars, respectively.
In section~\ref{Conclusions}, we make concluding remarks.

\section{Black hole X-ray binaries}\label{BHXB}

Astronomical black holes are characterized by two parameters, mass and spin,
and hence the measurement of these parameters is essential to probe the
fundamental physics of these objects \citep[e.g., ][]{Middleton2016}.
A black hole X-ray binary, i.e., 
a stellar-mass black hole accreting matter from a companion star,
is particularly useful for this purpose, as well as to test a theory of gravitation
and to probe the inflow and outflow of matter and its emission in an 
extremely strong gravitational field.

{\it AstroSat} SXT can be very useful to study soft X-ray continuum spectra
and spectral lines from BHXBs, to measure their temporal variations and to track their
evolution through various states. This telescope is particularly capable to 
characterize the accretion disk and reflection spectra, which can be used to
estimate black hole parameter values. Here, we will give a few examples.

\begin{figure*}[t]
        \centering\includegraphics[height=.35\textheight,angle=0]{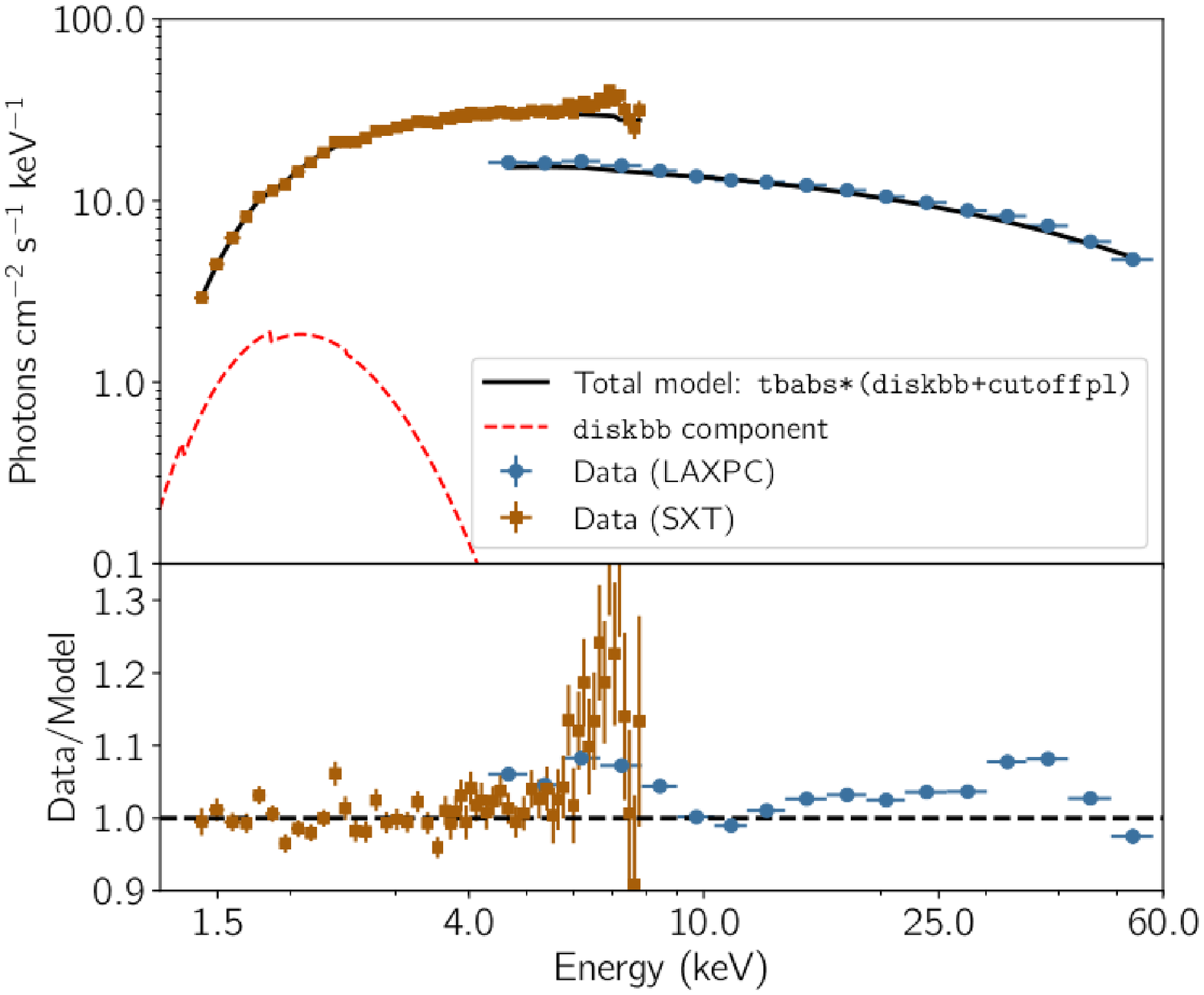}
	\caption{{\it AstroSat} SXT and LAXPC joint spectrum from the BHXB MAXI J1535--571. The model used to fit is an absorbed disk blackbody plus cut-off powerlaw, the latter representing a Comptonization component (upper panel). This brings out two prominent features -- Fe K$\alpha$ emission line and Compton hump -- of the reflection component of the spectrum in the data-to-model ratio plot (lower panel). It is clearly seen that both the disk blackbody and the asymmetry of the Fe K$\alpha$ line due to relativistic effects can be measured only with SXT, which are crucial to estimate the black hole mass and spin \citep[see section~\ref{BHXB}; figure courtesy: Navin Sridhar; ][]{Sridharetal2019}.
\label{sxt+laxpc_o2_ratio_binned_JAA.eps}}
\end{figure*}

\begin{figure*}[t]
	\centering\includegraphics[height=.35\textheight,angle=-90]{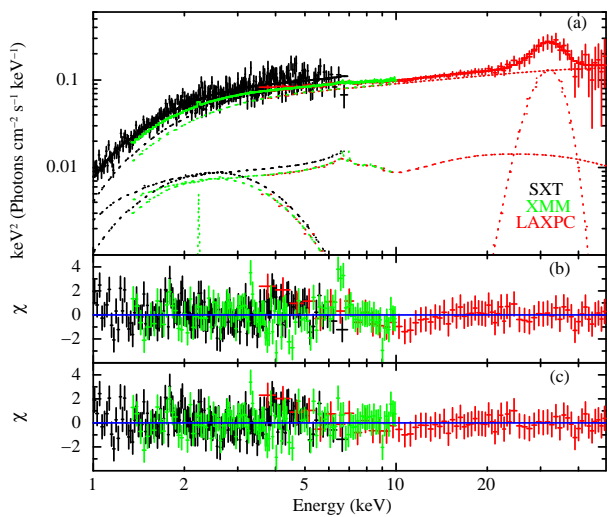}
	\caption{{\it AstroSat} SXT and LAXPC, and {\it XMM-Newton} EPIC-PN joint spectrum of the neutron star LMXB and accretion-powered millisecond X-ray pulsar SAX J1748.9--2021 (panel (a)). Panels (b) and (c) show the residuals for the XSPEC
models {\tt tbabs(bbodyrad+nthcomp)} and {\tt tbabs(bbodyrad+nthcomp+xillvercp)}, respectively. This figure shows that SXT and {\it XMM-Newton} EPIC-PN residuals match well with each other, with the SXT spectrum extending farther in lower energies \citep[see section~\ref{NSLMXB}; figure courtesy: Aru Beri; ][]{Sharmaetal2020}.
\label{plot-SXT-XMM-LAXPC.ps}}
\end{figure*}

A transient BHXB 4U 1630--47 was observed with {\it AstroSat} during its
2016 outburst. The source was found in a very soft state with an accretion 
disk contribution of $\sim 97$\% to the total flux. Such a source state is ideal 
to measure the black hole spin from the disk inner edge radius estimated
from the disk spectral component. SXT was particularly useful to characterize
the disk spectrum, having covered its peak and both sides better than other
instruments (see Fig.~\ref{4u1630-spec-sudip-v3.ps}). Using the 
XSPEC\footnote{https://heasarc.gsfc.nasa.gov/xanadu/xspec/} {\tt kerrbb}
model for the relativistic disk blackbody to fit {\it AstroSat} SXT+LAXPC and 
contemporaneous {\it Chandra} High Energy Grating (HEG) spectra, the
dimensionless black hole spin parameter was measured to be $0.92\pm0.04$
with $99.7$\% confidence \citep{Paharietal2018a}.
Similarly, a characterization of the thermal disk spectrum of the persistent 
BHXB LMC X--1 with SXT led to a measurement of the black hole
spin parameter of $\sim 0.93$ \citep{Mudambietal2020a}.

The ability of SXT to characterize both the disk blackbody spectral component 
and the relativistic Fe K$\alpha$ emission line of the reflection spectral 
component was crucial to measure mass and spin parameters of the transient 
BHXB MAXI J1535--571
\citep[see Fig.~\ref{sxt+laxpc_o2_ratio_binned_JAA.eps}; ][]{Sridharetal2019}.
More recently, {\it AstroSat} (SXT+LAXPC+CZTI) and contemporaneous {\it NuSTAR}
data were used to characterize another transient BHXB MAXI J1820+070. The black hole
mass was estimated to be $6.7-13.9$~M$_\odot$, and SXT, being the only instrument
in this work providing $< 3$~keV data and hence being able to reliably fit the disk
spectral component, was crucial for this estimation \citep{Chakrabortyetal2020}.
Note that this mass range derived from the spectral fitting is consistent with the results obtained from dynamical measurements \citep{Torresetal2019, Torresetal2020}.

In addition to the above examples, several other publications have reported
the characterization of BHXB spectral and timing properties with 
{\it AstroSat} science instruments, including SXT
\citep[e.g., ][]{Maqbooletal2019,Bhargavaetal2019,Sreeharietal2019,Sreeharietal2020,Mudambietal2020b,Babyetal2020}.
SXT was also useful to probe the formation of a giant radio jet base and
to measure the orbital period parameters of the X-ray binary Cygnus X--3,
for which the nature of the compact object is not yet confirmed 
\citep{Paharietal2018b,Bhargavaetal2017}.

\begin{figure*}[t]
        \centering\includegraphics[height=.35\textheight,angle=0]{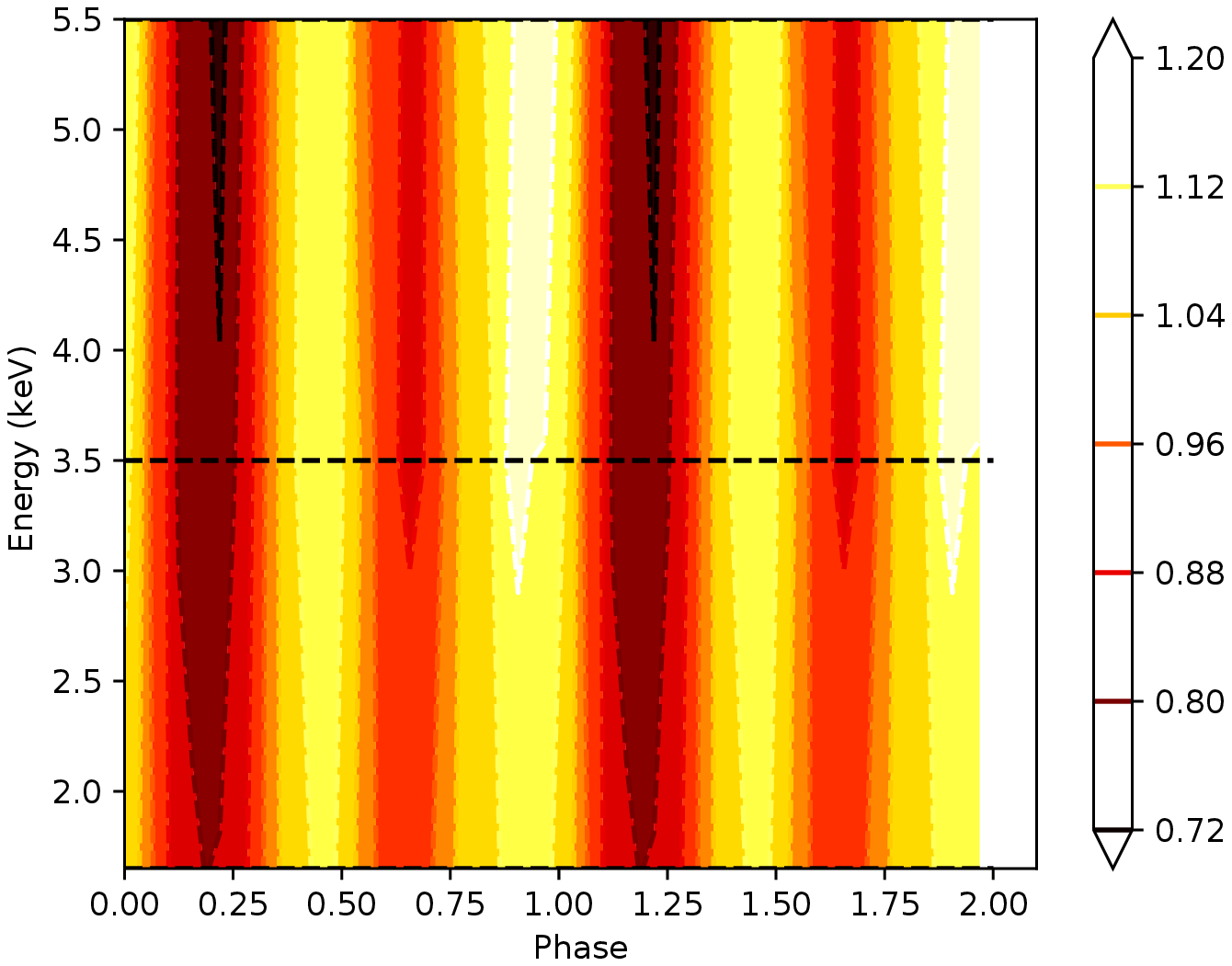}
	\caption{Energy-resolved pulse profile of the first Galactic ultra-luminous X-ray pulsar Swift J0243.6+6124, using the {\it AstroSat} SXT data. This figure shows that SXT could sufficiently resolve the X-ray pulsation period of $\sim 9.85$~s in its FW mode \citep[see section~\ref{ULP}; figure courtesy: Aru Beri; ][]{Berietal2020}.
\label{Heatmap-SXT-Obs2.eps}}
\end{figure*}

\section{Neutron star low-mass X-ray binaries}\label{NSLMXB}

A neutron star LMXB is a binary system in which the neutron star accretes
matter from a low-mass star. In such a system, the magnetic field of the
neutron star is typically low ($\sim 10^{7-9}$ G), and hence the accretion
disk can extend almost up to the stellar surface. Consequently, 
a number of features, such as thermonuclear X-ray bursts, high-frequency
quasi-periodic oscillations, accretion-powered millisecond period pulsations,
etc., are sometimes observed from these sources, which can be useful to probe
the strong gravity regime and the superdense degenerate core matter of 
neutron stars \citep{Bhattacharyya2010}.

{\it AstroSat} science instruments, including SXT, are ideal to study various spectral
and timing properties of neutron star LMXBs, and their evolution.
For example, the combined spectra from {\it AstroSat} (SXT+LAXPC) and 
{\it XMM-Newton} (EPIC-PN) observations of the neutron star LMXB SAX J1748.9--2021 
suggested the presence of reflection features \citep{Sharmaetal2020}.
Fig.~\ref{plot-SXT-XMM-LAXPC.ps} shows that the spectra measured with 
SXT and {\it XMM-Newton} EPIC-PN match well with each other.
Another work on the neutron star LMXB  GX 17+2 demonstrated that SXT not only is
suitable for spectral and timing studies in soft X-rays, but also
can be used for spectro-timing analyses \citep{Maluetal2020}. In this paper,
the cross-correlation studies using SXT and LAXPC light curves showed
time lags of the order of a hundred seconds.

\begin{figure*}[t]
        \centering\includegraphics[height=.35\textheight,angle=-90]{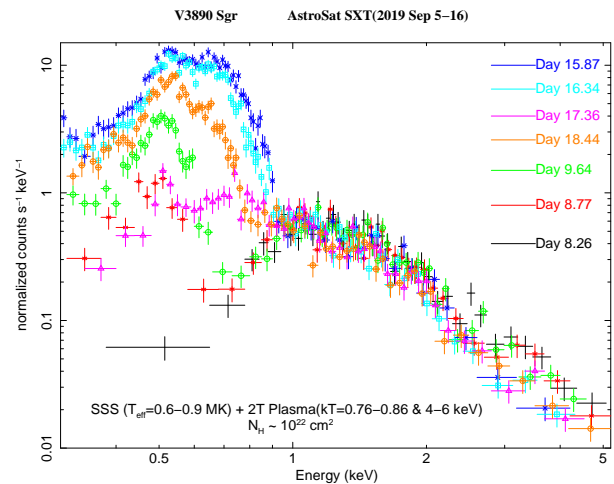}
	\caption{Evolution of {\it AstroSat} SXT spectrum of the symbiotic recurrent nova V3890 Sgr. The figure includes 7 of the 79 individual SXT spectra during its 2019 outburst, showing rapid variations. The days indicate the days after the beginning of the outburst. A Super Soft Source (SSS) emission component (with the best-fit effective temperature $\sim 0.6-0.9$ MK) and a two temperature plasma emission component (with best-fit temperatures $\sim 0.76-0.86$ keV and $\sim 4-6$ keV), with a best-fit hydrogen column density of $\sim 10^{22}$ cm$^{-2}$, were used to describe the spectra \citep[see section~\ref{CV}; ][]{Singhetal2020}.
\label{V3890Sgr_S1_S2_spec_evol.eps}}
\end{figure*}

\section{Neutron star high-mass X-ray binaries}\label{NSHMXB}

A neutron star HMXB is a binary system in which the neutron star accretes
matter from a high-mass star \citep[e.g., ][]{Walteretal2015}. 
In such a system, the magnetic field of the
neutron star is usually high ($\sim 10^{12}$ G), and hence the accretion
disk is typically truncated far from the star. As a result, the accreted matter
is channeled on to the magnetic polar caps, making the source a pulsar. 
These systems are ideal to study an interaction between the accreted matter 
and the strong stellar magnetic field. Several neutron star HMXBs, for example,
4U 0728--25, GRO J2058+42, 4U 1909+07, have been
observed with {\it AstroSat}, and SXT was used to characterize the broadband
spectra and pulse profiles in soft X-rays 
\citep{Royetal2020,Mukerjeeetal2020,Jaisawaletal2020}.

\section{Ultra-luminous X-ray pulsars}\label{ULP}

Ultra-luminous X-ray sources (ULXs), given their luminosities typically exceeding 
$10^{39}$ erg s$^{-1}$, could be accreting intermediate-mass black holes,
or neutron stars or stellar-mass black holes having accretion with 
super-Eddington rates. A subset of ULXs have been confirmed to be accreting
neutron stars from the observed spin-induced brightness pulsations, and they are known
as ultra-luminous X-ray pulsars \citep[ULPs; e.g., ][]{Kingetal2017}.

The characterization of the first
Galactic ULP, Swift J0243.6+6124, can therefore be extremely useful to understand
this important class of sources. {\it AstroSat} observed the 2017-2018 outburst
of Swift J0243.6+6124, and characterized its broadband spectrum, as well as
energy-dependent and luminosity-dependent pulse profiles in the energy range 
of $0.3-150$ keV \citep{Berietal2020}. SXT was particularly useful to 
measure the continuum spectrum, as well as pulse profiles
(see Fig.~\ref{Heatmap-SXT-Obs2.eps}), in soft X-rays.
Besides, recent observations of the Be X-ray binary and pulsar
RX J0209.6--7427 with {\it AstroSat} SXT and LAXPC have indicated that
its spectral and timing properties are remarkably similar to those
of ULXs, suggesting that this source could be a ULP \citep{Chandraetal2020}.

\begin{figure*}[t]
        \centering\includegraphics[height=.25\textheight,angle=0]{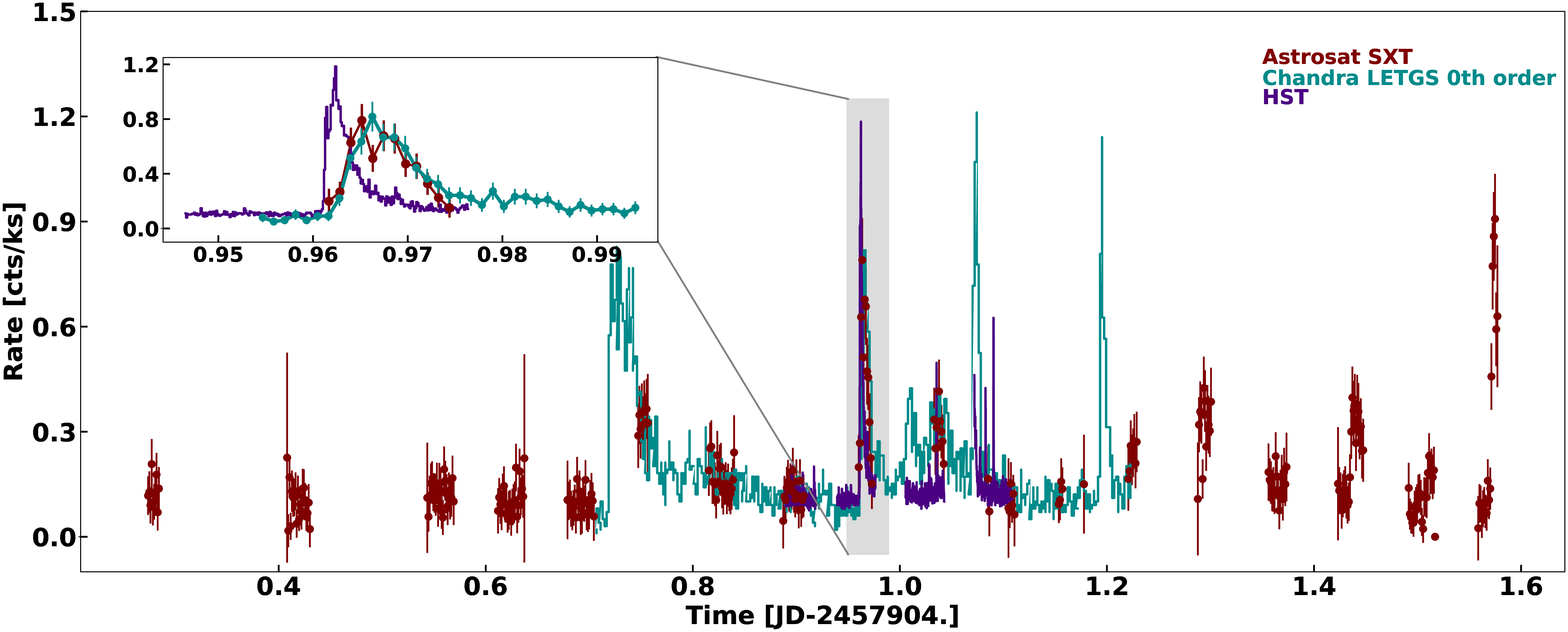}
	\caption{{\it AstroSat} SXT, {\it Chandra} LETGS 0th order and {\it HST} light curves of the nearest star Proxima Centauri, which hosts an Earth-like planet in its habitable zone. This figure shows flares and quiet periods of the star, which may affect the habitability of the planet. SXT observations were particularly useful to track the stellar activity throughout the observing campaign, with a flare observed simultaneously with all three instruments \citep[see section~\ref{star}; figure courtesy: Lalitha Sairam; ][]{Lalithaetal2020}.
\label{proxcen_lc_jaa.eps}}
\end{figure*}

\section{Cataclysmic variables}\label{CV}

{\it AstroSat} SXT can characterize spectral and temporal properties of 
cataclysmic variables (CVs), i.e., accreting white dwarfs.
A sub-type of such binary systems, in which the white dwarf accretes matter from 
a red giant donor star via an accretion disk, can have explosive thermonuclear 
burning of the accumulated hydrogen rich material.
This may lead to an outburst with a massive ejection of the material at 
velocities $\ge 300$ km s$^{-1}$.
These are known as Symbiotic Recurrent Novae, and only four such objects 
are currently known to exist \citep{Schaefer2010}. 

{\it AstroSat} SXT observed one such nova, 
V3890 Sgr, in two long observations in 2019 from 5th September to 16th September, 
just $\sim 8$ days after its third recorded outburst, with the highest cadence 
monitoring from a low-Earth orbit satellite \citep{Singhetal2020}.
The observations caught the first 
appearance of Super Soft Source (SSS) emission ($< 1$~keV) on day 8.57 after 
the outburst, revealing the presence of a very high mass white dwarf. Rapid and 
highly variable evolution of the SSS, that included its complete vanishing during 
days $8.6-8.9$ and subsequent appearance, followed by another extremely low flux 
state during days $16.8-17.8$, and rising again were observed. A detailed spectral 
modeling, using white dwarf emission models for the SSS and plasma models 
for higher energy ($1-7$ keV) emission, to study the source spectral 
evolution has been carried out.
The rapid spectral evolution (see Fig.~\ref{V3890Sgr_S1_S2_spec_evol.eps}) 
on nearly hourly time scales is not explained by evolutionary models of 
accretion and ejecta.

\section{Active galactic nuclei}\label{AGN}

Active galactic nuclei (AGNs) are supermassive black holes, which accrete gas 
from the surrounding medium near the centres of galaxies. AGNs can radiate 
prominently in a wide energy range -- from radio to $\gamma$-rays,
and are rich in observational features in multiple wavebands. 
Based on these features, these objects have been divided into several
subclasses, such as Seyfert galaxies and Blazars. AGNs are extremely
important to study the strong gravity regime, accretion/ejection mechanisms,
the feedback to the host galaxy and the intergalactic medium, and cosmology
\citep[e.g., ][]{Netzer2015}.

{\it AstroSat} has observed several AGNs, and SXT has significantly contributed
to the spectral and timing characterization of some of them. 
For example, SXT has measured the broadband spectrum and variability of the narrow-line 
Seyfert 1 galaxy RE J1034+396 in soft X-rays, and contributed to the 
source power spectrum, which, when compared with the power spectrum of the
BHXB GRS 1915+105, indicated a supermassive black hole mass of 
$\sim 3\times10^6$~M$_\odot$ \citep{Chaudhuryetal2018}.
In another work, the variability of a Blazar, Mrk 421, was measured
with {\it AstroSat} SXT and LAXPC, as well as with {\it Swift}, which
was very promising to establish a way to probe the disk-jet connection 
\citep{Chatterjeeetal2018}. Besides, SXT was useful to characterize the 
soft X-ray spectrum of another Blazar, RGB J0710$+$591 \citep{Goswamietal2020}.

Simultaneous UV/X-ray observations of a Seyfert galaxy with {\it AstroSat}
can be useful to probe the thermal Comptonization responsible for 
the broadband X-ray emission. Joint X-ray spectral analyses of five sets of 
SXT and LAXPC spectra revealed a steepening and brightening X-ray power-law 
component with increasing intrinsic UV emission for the Seyfert 1.2 AGN
IC 4329A (Tripathi et al., to be submitted). 
These observations implied that UV emission from the
disk indeed provides the primary seed photons for the Thermal Comptonization process,
and the X-ray spectral variability is caused by either cooling of the hot corona 
or increasing optical depth of the corona, each with increasing UV flux.

\section{Stars}\label{star}

Coronal activity in stars can be studied by observing the variation of 
its high-energy radiation, including occasional flares \citep{GudelNaze2009}. 
Such a study can be 
useful not only to probe the stellar physics and the surrounding environment,
but also to understand the impact of the ejected particles and radiation on
the plausible planets, including their habitability. {\it AstroSat} SXT
is capable of tracking the stellar soft X-ray intensity variation, which was demonstrated
by an observational campaign on our nearest star Proxima Centauri with
SXT, {\it Chandra} Low Energy Transmission Grating Spectrometer (LETGS) 
and {\it Hubble 
Space Telescope} ({\it HST}). Proxima Centauri is an M-dwarf with an Earth-like
planet within its habitable zone, and several flares and the non-flare
emission observed from the star were useful to probe the coronal temperatures, 
abundance, etc. \citep[see Fig.~\ref{proxcen_lc_jaa.eps}; ][]{Lalithaetal2020}.
Particularly, one flare was observed with all three instruments, and showed the
Neupert effect, that is the UV emission preceding the soft X-ray emission.

\section{Conclusions}\label{Conclusions}

Recent publications in refereed journals have confirmed that SXT can
successfully study spectral and timing properties of a variety of cosmic 
sources, as a standalone telescope, as well as in combination with 
other {\it AstroSat} science instruments and other satellites, such as {\it Chandra},
{\it XMM-Newton}, {\it NuSTAR}, and even {\it HST}. Since the launch of
{\it AstroSat} in late 2015, SXT has been performing as expected, without any 
significant degradation of its capabilities. If this continues, SXT can be very
useful for deep observations, as well as to track the evolution of 
relatively bright X-ray transients in a more dedicated manner in the future.



\section*{Acknowledgements}

The publications mentioned here used the data from the AstroSat mission of the Indian Space Research Organisation (ISRO), archived at the Indian Space Science Data Centre (ISSDC).
The works mentioned here were performed utilizing the calibration data-bases and
auxiliary analysis tools developed, maintained and distributed by the
AstroSat-SXT team with members from various institutions in India and
abroad, and the SXT Payload Operation Center (POC) at the TIFR, Mumbai
(https://www.tifr.res.in/$\sim$astrosat\_sxt/index.html). SXT data were
processed and verified by the SXT POC.
We thank Mayukh Pahari, Navin Sridhar, Aru Beri and Lalitha Sairam for providing some of the figures.

\bibliography{Bhattacharyya_SXT_biblio}

\end{document}